\documentclass{emulateapj}  

\usepackage{graphics}   
\usepackage{graphicx}  
\usepackage{epstopdf}   
\usepackage{epsfig}   

\begin{document}

\slugcomment{Accepted to ApJL -- 10/20/08}

\title{New Observations and a Possible Detection of Parameter Variations in the Transits of Gliese 436\MakeLowercase{b}}

\author{Jeffrey L. Coughlin\altaffilmark{1}, Guy S. Stringfellow\altaffilmark{2}, Andrew C. Becker\altaffilmark{3}, Mercedes L\'opez-Morales\altaffilmark{4}, Fabio Mezzalira\altaffilmark{2}, Tom Krajci\altaffilmark{5}}

\altaffiltext{1}{Department of Astronomy, New Mexico State University, P.O. Box 30001, MSC 4500, Las Cruces, New Mexico 88003-8001; jlcough@nmsu.edu}
\altaffiltext{2}{Center for Astrophysics and Space Astronomy, Department of Astrophysical and Planetary Sciences, University of Colorado, 389 UCB, Boulder, CO 80309-0389; Guy.Stringfellow@colorado.edu}
\altaffiltext{3}{Department of Astronomy, University of Washington; becker@astro.washington.edu}
\altaffiltext{4}{Hubble Fellow, Department of Terrestrial Magnetism, Carnegie Institution of Washington; mercedes@dtm.ciw.edu}
\altaffiltext{5}{Astrokolkhoz Observatory, PO Box 1351, Cloudcroft, NM 88317; tom\_krajci@tularosa.net}

\begin{abstract}

We present ground-based observations of the transiting Neptune-mass planet Gl 436b obtained with the 3.5-meter telescope at Apache Point Observatory and other supporting telescopes. Included in this is an observed transit in early 2005, over two years before the earliest reported transit detection. We have compiled all available transit data to date and perform a uniform modeling using the JKTEBOP code. We do not detect any transit timing variations of amplitude greater than $\sim$1 minute over the $\sim$3.3 year baseline. We do however find possible evidence for a self-consistent trend of increasing orbital inclination, transit width, and transit depth, which supports the supposition that Gl 436b is being perturbed by another planet of $\lesssim$ 12 M$_{\earth}$ in a non-resonant orbit.

\end{abstract}


\section{Introduction}

Gliese 436 is an M-dwarf (M2.5V) with a mass of 0.45 M$_{\sun}$ and hosts the extrasolar planet Gl 436b, which is currently the least massive transiting planet with a mass of 23.17 M$_{\earth}$ \citep{Torres07}, and the only planet known to transit an M dwarf. Gl 436b was first discovered via radial-velocity (RV) variations by \citet{Butler04}, who also searched for a photometric transit, but failed to detect any signal greater than 0.4\%. It was thus a surprise when \citet{Gillon07b} reported the detection of a transit with a depth of 0.7\%, implying a planetary radius of 4.22 R$_{\earth}$ \citep{Torres07} and thus a composition similar to Uranus and Neptune.  In addition, both \citet{Deming07} and \citet{Maness07} calculated that the significant eccentricity of the orbit, e = 0.15, coupled with its short period of $\sim$2.6 days, should result in circularization timescales of $\sim$10$^{8}$ years, which contrasts with the old age of the system at $\ga$6$\times$10$^{9}$ years. The existence of one or more additional planets in the system could be responsible for perturbations to Gl 436b's orbit, and thus result in the observed peculiarities. We considered this possibility right after the initial publication of \citet{Gillon07b}, and began an intensive campaign to observe the photometric transits of Gl 436b in order to search for variations indicative of orbital perturbations \citep{Stringfellow08}.

Early this year, \citet{Ribas08a} reported the possible detection of a $\sim$5 M$_{\earth}$ companion in the Gl 436 system located near the outer 2:1 resonance of Gl 436b via analysis of all the RV data compiled to date. Theoretically this planet would be perturbing Gl 436b so as to increase its orbital inclination at a rate of $\sim$0.1 deg yr$^{-1}$, and thus its transit depth and length, so that the non-detection by \citet{Butler04} and the observed transit of \citet{Gillon07b} were compatible. Since the RV detection of this second planet had a significant false-alarm probability of $\sim$20\%, \citet{Ribas08a} proposed that confirmation could be achieved through 2008 observations of Gl 436b's transits, which would show a lengthening of transit duration by $\sim$ 2 minutes compared to the \citet{Gillon07b} data. As well, transit-timing variations (TTVs) of several minutes should also be detectable by observing a significant number of transits.

Recently, \citet{Alonso08} reported a lack of observed inclination changes and TTV evidence for the second planet, based on a comparison of a single H band light curve obtained in March 2008 to 8$\micron$ data taken with Spitzer 254 days earlier \citep{Gillon07a,Deming07}. This result, combined with additional radial velocity measurements \citep{Howard08,Bonfils08} that contradicted the proposed period of the second planet, drove \citet{Ribas08b} to retract their claim of the companion at IAU Symposium 253. However, very recently \citet{Shporer08} presented multiple light curves obtained in May 2007, and could not rule out TTVs on the order of a minute. While the planet specifically proposed by \citet{Ribas08a} most likely does not exist,  \citet{Ribas08b} makes a strong case that a second planet is still needed to explain the peculiarities of Gl 436b, and most likely exists in a non-resonant configuration where no strong TTVs are induced. Amateur astronomers have been diligent in observing Gl 436b since it's initial transit discovery, and thus along with this data, published data, and our own data, we are able to present a thorough analysis of the TTVs, inclination, duration, and depth of the transit changes in the Gliese 436 system. We present our observations in \S2, our modeling and derivation of parameters in \S3, and explore the observed TTVs and parameters of the system over time in \S4.\\
\\

\section{Observations}

We observed Gl 436 (11h42m11s, +26$\degr$42$\arcmin$24$\arcsec$ J2000) in the V filter on the nights of April 7, April 28, and May 6 2008 UT with the 3.5-meter telescope at Apache Point Observatory (APO). We used a backside-illuminated SITe 2048x2048 CCD with 2x2 binning, resulting in a plate scale of 0.28$\arcsec$/pixel, and sub-framed to a field of view of 4.8$\arcmin$ by 0.56$\arcmin$ to decrease readout time. We applied typical overscan, bias, and flat-field calibrations. For photometric reduction we used the standard IRAF task PHOT, with the aperture selected as a constant multiple of the Gaussian-fitted FWHM of each image to account for any variable seeing. We performed differential photometry with respect to the star USNO 1167-0208653 (2MASS ID 175252970) located at 11h42m12.08s, +26$\degr$46$\arcmin$07.45$\arcsec$ J2000. This star has V=10.82 and color V-I=1.48, compared to Gl 436 which has V=10.68, and color V-I=1.70. In the error bar computation, we account for both standard noise from the photometry, as well as due to scintillation following equation 10 of \citet{Dravins98}. Having obtained at least 30 minutes of data on each side of the transit, we subtracted a linear fit for all data outside of transit vs. airmass to account for any differential reddening. Resulting individual data points have errors ranging from 1.5 to 2.8 mmag, which agrees with the rms of the residuals from the model fits, and a typical cadence of about 17 seconds. We have searched for correlated noise on the timescale of ingress and egress, via the technique of \citet{Pont06}, but only find a statistically significant amount for the night of April 7, measured to be 0.11 mmag. The three transits are shown in Figure 1. 

\begin{figure}
\centering
\includegraphics[angle=0,width=0.75\linewidth]{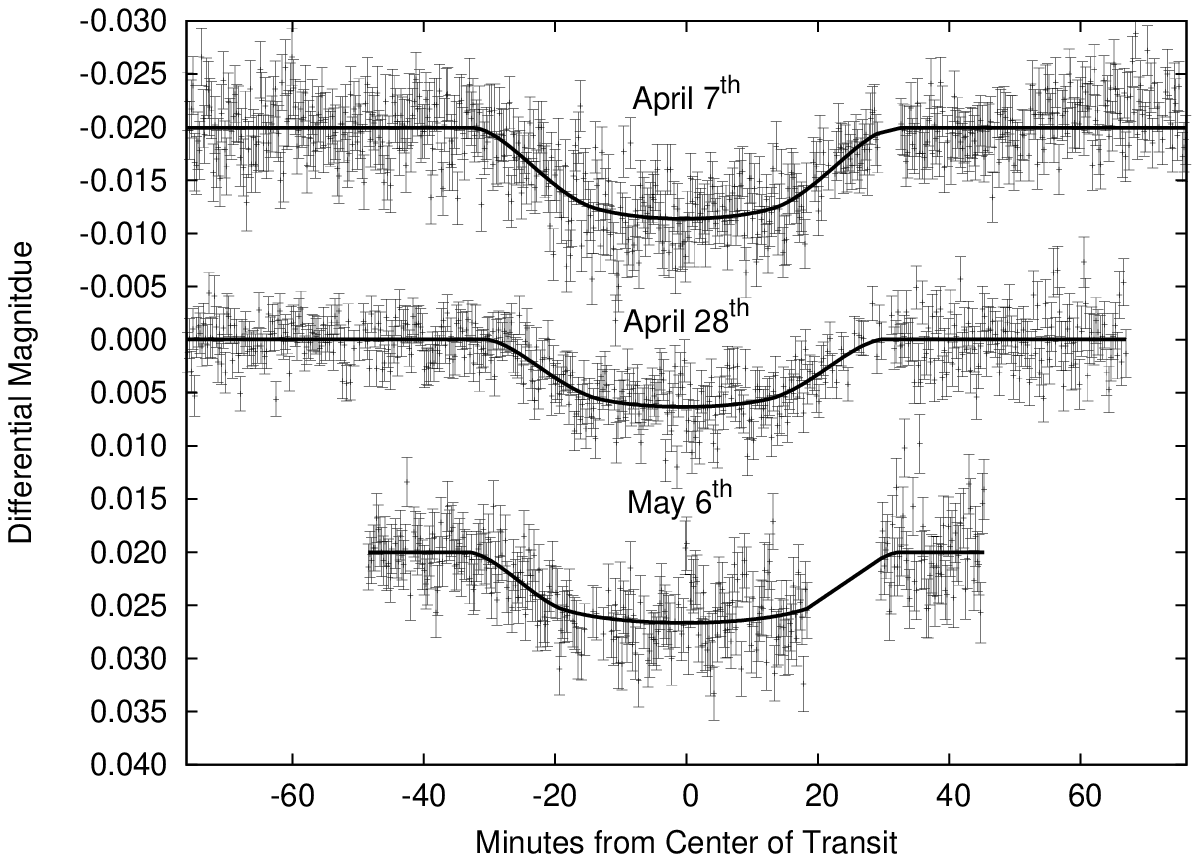}
\includegraphics[angle=0,width=0.75\linewidth]{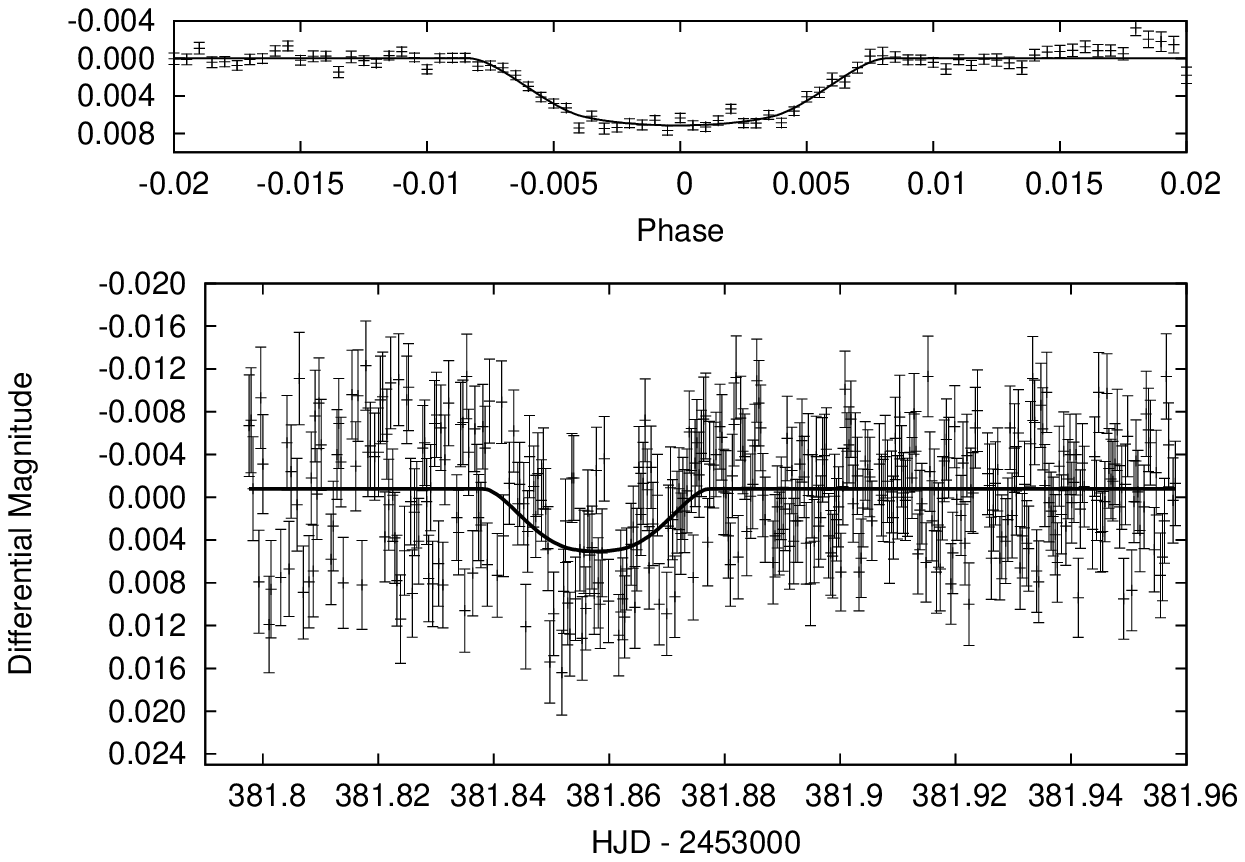}
\caption{\emph{Top}: The V band light curves observed with the APO 3.5-meter with model fits. \emph{Middle}: The 3.5-meter data combined, phased, and binned in increments of 0.0005 phase. \emph{Bottom}: The transit observed by the NMSU 1-meter telescope on the night of January 11 2005 UT. A 3-sigma clip has been applied, and is shown with a model fit for which the radii were fixed.}
\end{figure}

We also carried out accompanying observations with the New Mexico State University (NMSU) 1-meter telescope at APO, in the V filter on the night of April 7 2008 UT, and in the I filter on the night of April 28 2008 UT. A 2048x2048 E2V CCD was used with 1x1 binning and sub-framing, resulting in a field of view of 8.0$\arcmin$ square and a plate scale of 0.47$\arcsec$/pixel, and we applied the aforementioned standard calibration and photometric extraction techniques. We performed ensemble photometry with respect to the USNO star that was used as the 3.5m reference, as well as BD+27 2046 (V=10.64, V-I=0.44), and another star at 11h42m00s, +26$\degr$45$\arcmin$56$\arcsec$ J2000 (V=12.81, V-I=1.46). Resulting typical errors on individual points range from 3 to 5 mmag with a typical cadence of about 12 seconds.

The NMSU 1-meter telescope can also function as a robotic telescope, and is used intermittently to photometrically monitor stars with known radial-velocity discovered planets to search for transits. A search of the 1-meter archives revealed that it observed Gl 436 on the night of January 11 2005 UT, during which a transit should have occurred, according to the precise ephemeris for Gl 436b that is now available by incorporating the many observed transits in 2007 and 2008. At the time, this 1-meter program depended on visual inspection of automatically generated photometry and plots. For this night, the plot had large temporal and brightness ranges, and thus the tiny transit was easily missed. However, now carefully inspecting the region constrained by the ephemeris, as well as re-performing the photometry to maximize signal-to-noise, we find a transit signature within a minute of that predicted by the ephemeris with reasonable width and depth, as shown in Figure 1. Individual data points have an error of about 4 mmag, a cadence of 30 seconds, and we do not detect any correlated noise with any level of significance.
 
We also conducted observations on the nights of April 28 and May 13 2008 UT using a 24$"$ telescope located at the Sommers-Bosch Observatory (SBO) on the University of Colorado at Boulder campus, using an I filter. These observations also used a windowed chip and an exposure time to maximize signal-to-noise without saturating, and have comparable temporal resolution to the 3.5m and 1m telescopes due to a shorter readout time. As well, we used an unfiltered 11$"$ telescope at Cloudcroft, NM (CC) with a SBIG ST-7E CCD and 2x2 binning on May 6 2008 UT, with a resulting cadence of about 25 seconds. We have also gathered all the amateur data currently available on the system as compiled by Bruce Gary \anchor{http://brucegary.net/AXA/GJ436/gj436.htm}{(http://brucegary.net/AXA/GJ436/gj436.htm)}.

\section{Modeling and Derivation of Parameters}

We use the JKTEBOP code \citep{Southworth04a,Southworth04b} to model all the transit light curves in a consistent and uniform manner. \citet{Southworth08} has recently performed an exhaustive analysis of fourteen transiting planets using the JKTEBOP code, and shows it compares well with results reported elsewhere. JKTEBOP offers the advantage of incorporating a Levenberg-Marquardt optimization algorithm, improved limb darkening treatments, and extensive error analysis routines, which are critical for confirming any trends in the system.

For each transit curve, we solved for the ratio of radii (k = R$_{p}$/R$_{s}$), the orbital inclination (i), the time of mid-transit (T$_{0}$), and a scale factor that defines the normalized value of the out-of-transit flux in the light curves. In order to obtain reasonable results for the scale of the system for all data sets, the sum of the radii (R$_{s}$ + R$_{p}$) was set to that found by \citet{Torres07}. We also fixed the eccentricity to a value of 0.15 and the longitude of periastron to 343$\degr$ as given by \citet{Deming07} and \citet{Mardling08}. We used a quadratic limb-darkening law with coefficients taken from \citet{Claret00} for T$_{eff}$ = 3500K, log(g) = 4.5, V$_{t}$ = 2.0 km s$^{-1}$, and [M/H] = 0.0, for the appropriate filters. In the case of the Spitzer 8$\micron$ data, we used the coefficients as determined by \citet{Gillon07a}. From each fit, still assuming a constant sum of radii, we were thus also able to calculate the individual star and planet radii, as well as the depth and width of transit. In order to rule out any potential correlations in derived planet size and inclination, we then re-modeled all data with the same procedure, but also fixing k, and thus the star and planet sizes, to that found by \citet{Torres07}. This generally produced similar results, but for the noisier data sets achieved more consistent results. Parameters from both techniques are shown in Table 1.

In order to obtain robust errors, we ran 1,000 Monte-Carlo simulations for each data set and performed a residual-permutation analysis \citep{Jenkins02} to investigate temporally correlated noise. In both cases, the previously fixed parameters, as well as the limb-darkening coefficients, were allowed to vary so that their individual uncertainties would be taken into account in the derived parameter uncertainties. For each Monte Carlo simulation, random Gaussian noise with amplitude equal to the given error bars, or in the absence thereof the standard deviation of the residual scatter from the best-fit solution, was added to each data point and the curve re-fitted with random perturbations applied to the initial parameter values. This ensured a detailed exploration of the parameter space and parameter correlations. However, this Monte Carlo technique will underestimate errors for certain parameters in the presence of temporally correlated noise, which can result from trends in seeing, extinction, focus, or other atmospheric or telescope related phenomena \citep{Southworth08}. The residual-permutation method takes the residuals of the best-fit model, shifts them to the next data point, and finds a new solution. The residuals are shifted again, a new fit is found, and the process repeats as many times as there are datapoints. Thus, there is a distribution of fitted values similar to the Monte Carlo technique, but any temporal trends will have been propagated around the light curve, and thus taken into account. For our final errors we adopt the larger value found between the two methods, although for the majority of parameters and data sets the two methods agree quite well.

In total we modeled 28 light curves, (16 professional and 12 amateur), covering 19 separate transit events over a baseline of nearly 3.3 years.

\section{Transit Timing and Eclipse Variations}

Using the derived time of minima in Table 1 for all the data when allowing k to vary, we derive a new linear, error-weighted ephemeris of  T$_{c}$(HJD) = 2454222.6164(1) + 2.643897(2)$\cdot$E, where the parentheses indicate the amount of uncertainty in the last digit, and E is the epoch with E = 0 the initial transit discovery of \citet{Gillon07b}. Using this ephemeris, we then compute an observed minus calculated (O-C) diagram for the time of transit center, as shown in Figure 2. We have currently excluded the amateur data from the plot due to much larger error bars, so that the high-precision data points can be seen clearly. We have examined the TTVs and various subsets thereof using a phase dispersion minimization technique \citep{Stellingwerf78}, but do not find any periods with statistical significance. Examining the best data, specifically the previously published data and our 3.5-meter observations, there is a standard deviation of 52 seconds. Assuming a sinusoidal TTV trend, we can then rule out any TTVs with amplitude greater than $\sim$ 1 minute.

\begin{figure}
\centering
\includegraphics[angle=0,width=\linewidth]{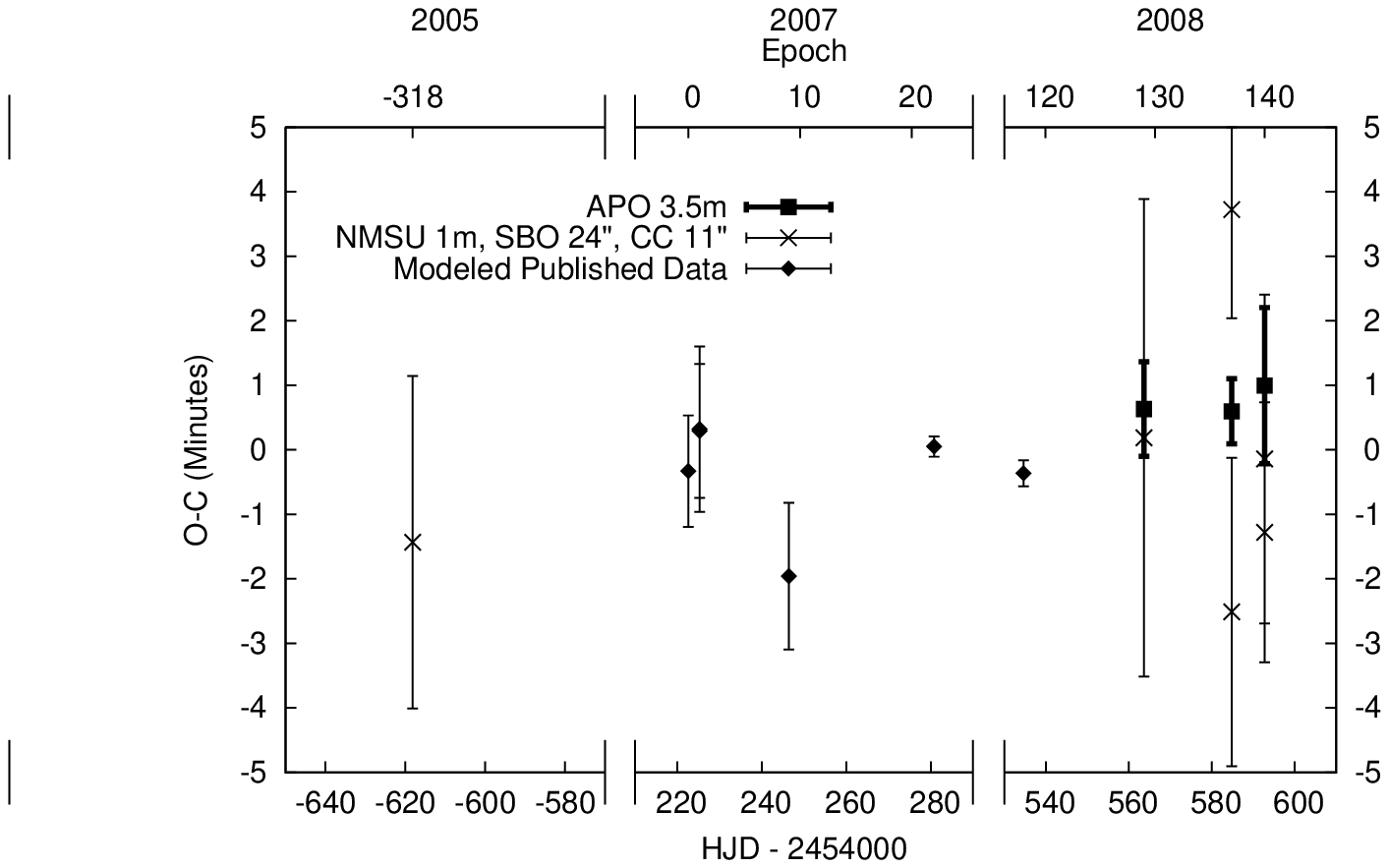}
\caption{O-C diagram for all professional times of minima.}
\end{figure}

We have searched for any trends in derived inclination, width, and depth of transit over time via error-weighted least-squares linear regression. In addition, we have also performed 10,000 Monte Carlo simulations for each fit, where gaussian noise with amplitude equal to each point's error bars was added in each iteration and the data re-fitted, with resulting 1$\sigma$ parameter distributions giving robust errors. The two methods agree to within 1\% for all values. As mentioned in \S3, we modeled all the light curves by both allowing the ratio of radii to vary as well as fixing it, and thus we list the values for each set. Performing fits to all the data, we have a tentative detection of increasing inclination, transit width, and transit depth with time, as shown in Table 2. We present these fits with the actual data derived when fixing the radii in Figure 3. As a precaution against any bias being introduced by the much larger number of data points at later epochs, we decided to separately bin the 2005, 2007, and 2008 data using an error-weighted mean, and re-fit the three resulting data points for each modeling method. As shown in Table 2, the values agree very well with those derived when not binning the data.

\begin{deluxetable}{lccccc}
\tablewidth{0.0pt}
\tabletypesize{\scriptsize}
\tablenum{2}
\tablecaption{Trends in derived inclination, width, and depth of transit over time}
\tablecolumns{5}
\tablehead{Radii & Data Set & deg yr$^{-1}$ & min yr$^{-1}$ & mmag yr$^{-1}$}
\startdata
 & All & 0.120$\pm$0.062 & 3.43$\pm$1.01 &  0.28$\pm$0.16\\
Variable & Binned & 0.126$\pm$0.061 & 3.53$\pm$0.97 & 0.26$\pm$0.14\\
 & No 2005 & 0.092$\pm$0.099 & 3.10$\pm$1.10 & 0.29$\pm$0.17\\
\hline
 & All & 0.069$\pm$0.051 & 2.36$\pm$0.84 &  \ 0.32$\pm$0.20\\
Fixed & Binned & 0.071$\pm$0.050 & 2.37$\pm$0.81 & \ 0.32$\pm$0.19\\
 & No 2005 & 0.020$\pm$0.099 & 1.68$\pm$1.29 & -0.01$\pm$0.42\\
\enddata
\end{deluxetable}

\begin{figure}
\centering
\includegraphics[angle=0,width=\linewidth]{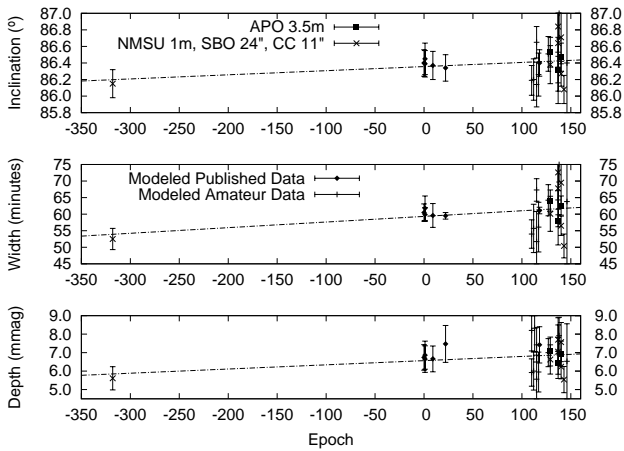}
\caption{Measured inclination, width, and depth of transit over time for all data, with the star and planet radii fixed.}
\end{figure}

The trends are moderately dependent on the single 2005 transit data point, which greatly extends the temporal baseline, and as such we are cautious about any claims. Resulting temporal trends when removing the 2005 data point are also shown in Table 2. Although while removing the 2005 data point significantly weakens the claim of a variation of inclination with time, the trend of increasing width still holds. Also of interest is that at a rate of 0.120 deg yr$^{-1}$, as derived from our fit to all the data fitted with a variable radius, the JKTEBOP program yields an increase in transit width of 4.36 min yr$^{-1}$, and depth of 0.544 mmag yr$^{-1}$, which are in agreement with our observed trends, and thus are self-consistent. As well, the measured rate of inclination change is compatible with the $\sim$0.1 deg yr$^{-1}$ required to make congruent the non-detection of \citet{Butler04} and the observed transit of \citet{Gillon07b}. Extending the measurement baseline a couple years into the future will confirm or negate this result.

\section{Discussion and Conclusion}

We have presented a total of ten new transit light curves of Gl 436b, three of which come from the 3.5-meter telescope at APO, and one of which is from the NMSU 1-meter in January 2005. We have collected and uniformly modeled all available professional and amateur light curves, and searched for any trends in transit timing, width of transit, and depth of transit variations. We find statistically significant, self-consistent trends that are compatible with the perturbation of Gl 436b by a planet with mass $\lesssim$ 12 M$_{\earth}$ in a non-resonant orbit with semi-major axis $\lesssim$ 0.08 AU. This conclusion is based on the numerical simulations of \citet[][see Fig. 1]{Ribas08a} who constrain the mass and semi-major axis of the theoretical second planet by examining which configurations could produce the observed orbital perturbations while still remaining undetected by the existing radial-velocity data. From our analysis, we infer a non-resonant orbit based on a lack of detected TTVs with amplitude $\gtrsim$ 1 minute. 

We stress that our measured trends are moderately dependent on our 2005 data, and thus subsequent high-precision observations over the next few years need to be carried out to confirm or refute this trend. If confirmed, it would be strong evidence for the first extrasolar planet discovered via orbital perturbations to a transiting planet. Also, we would like to note that although \citet{Alonso08} had previously limited the rate of inclination change to 0.03$\pm$0.05 deg/yr, they did so only by measuring the change in width between the 2007 Spitzer observations and their own 2008 H-band data, which they found to be 0.5$\pm$1.2 minutes. Via Table 1, we find the difference in transit width between the two observations to be 1.5$\pm$1.4 minutes, which is in agreement with our derived inclination and width values, and is a more reliable result due to using full model fits with proper limb-darkening coefficients. With respect to the the amateur observations, although they are numerous, the very small depth of the transit makes it a challenge for most small aperture systems, resulting in very large uncertainties in i and T$_{0}$. Also, while amateur observers are aware of the importance of precision timing, we of course cannot examine each of their observing set-ups, and thus one must be aware of the possibility, although small, of systemic time offsets on a given night when interpreting their data.

\acknowledgments
JLC acknowledges support from the New Mexico Space Grant Consortium and the National Science Foundation, and GSS is supported by grants from NASA. MLM acknowledges support provided by NASA through Hubble Fellowship grant HF-01210.01-A awarded by the STScI, which is operated by the AURA, Inc., for NASA, under contract NAS5-26555. The authors would like to thank Eric Algol for helpful discussions, and Roi Alonso for providing his H band data. We also thank the many amateur astronomers for dedicated monitoring of the Gl 436 system, and to Bruce L. Gary for compiling their data.


\begin{deluxetable}{lccccccccc}
\tablewidth{0.0pt}
\tabletypesize{\scriptsize}
\tablenum{1}
\tablecaption{Parameters derived from using the JKTEBOP code with 1$\sigma$ errors}
\tablecolumns{9}
\tablehead{Epoch & Source & Filter\tablenotemark{c} & Tmin & Inclination & R$_{\star}$ & R$_{P}$ & Depth & Width\\ & & & (HJD-2450000) & (deg) & (R$_{\sun}$) & (R$_{\earth}$) & (mmag) & (minutes)}
\startdata
\cutinhead{Ratio of Radii k Allowed to Vary}
-318 & NMSU 1m & V & 3381.85584$\pm$0.00179 & 86.02$\pm$0.23 & 0.446$\pm$0.046 & 6.19$\pm$4.81 & 7.05$\pm$1.35 & 47.0$\pm$7.1\\
0 & \citet{Gillon07b}\tablenotemark{b} & V & 4222.61617$\pm$0.00060 & 86.38$\pm$0.18 & 0.463$\pm$0.016 & 4.32$\pm$0.24 & 6.98$\pm$0.43 & 60.1$\pm$1.6\\
1 & \citet{Shporer08}\tablenotemark{b} & None & 4225.26052$\pm$0.00089 & 86.43$\pm$0.17 & 0.463$\pm$0.015 & 4.33$\pm$0.27 & 7.16$\pm$0.81 & 61.4$\pm$2.5\\
1 & \citet{Shporer08}\tablenotemark{b} & V & 4225.26050$\pm$0.00072 & 86.35$\pm$0.17 & 0.462$\pm$0.016 & 4.47$\pm$0.25 & 7.31$\pm$0.46 & 59.2$\pm$1.9\\
9 & \citet{Shporer08}\tablenotemark{b} & R & 4246.41012$\pm$0.00079 & 86.27$\pm$0.18 & 0.456$\pm$0.014 & 5.10$\pm$0.66 & 9.07$\pm$0.87 & 56.7$\pm$2.6\\
22 & \citet{Gillon07a}  & 8$\micron$ & 4280.78219$\pm$0.00011 & 86.34$\pm$0.16 & 0.464$\pm$0.016 & 4.23$\pm$0.16 & 7.46$\pm$0.10 & 59.8$\pm$1.1\\
110 & Gregor Srdoc\tablenotemark{a}  & R & 4513.43393$\pm$0.00174 & 86.10$\pm$0.24 & 0.457$\pm$0.030 & 5.04$\pm$2.91 & 7.11$\pm$1.29 & 49.8$\pm$6.5\\
110 & Tonny Vanmunster\tablenotemark{a}  & R & 4513.44404$\pm$0.00247 & 87.13$\pm$0.30 & 0.461$\pm$0.016 & 4.57$\pm$0.45 & 9.63$\pm$1.67 & 77.8$\pm$4.9\\
112 & Bruce Gary\tablenotemark{a}  & R & 4518.72999$\pm$0.00278 & 86.04$\pm$0.34 & 0.443$\pm$0.087 & 6.55$\pm$9.38 & 8.78$\pm$2.37 & 47.0$\pm$11.7\\
113 & Gregor Srdoc\tablenotemark{a}  & R & 4521.37338$\pm$0.00130 & 87.27$\pm$0.30 & 0.459$\pm$0.015 & 4.81$\pm$0.30 & 11.06$\pm$1.13 & 80.0$\pm$4.2\\
115 & James Roe\tablenotemark{a}  & V & 4526.65995$\pm$0.00124 & 86.03$\pm$0.27 & 0.447$\pm$0.036 & 6.05$\pm$3.34 & 7.47$\pm$1.55 & 48.4$\pm$9.0\\
115 & Joao Gregorio\tablenotemark{a}  & V & 4526.65972$\pm$0.00130 & 87.09$\pm$0.28 & 0.468$\pm$0.016 & 3.80$\pm$0.25 & 6.50$\pm$0.66 & 77.1$\pm$4.0\\
117 & Richard Schwartz\tablenotemark{a}  & V & 4531.94399$\pm$0.00222 & 86.18$\pm$0.20 & 0.461$\pm$0.015 & 4.51$\pm$0.65 & 6.37$\pm$1.67 & 53.4$\pm$6.5\\
118 & \citet{Alonso08} & H & 4534.59611$\pm$0.00014 & 86.39$\pm$0.17 & 0.463$\pm$0.016 & 4.32$\pm$0.17 & 7.74$\pm$0.11 & 61.1$\pm$0.9\\
127 & Manuel Mendez\tablenotemark{a} & R & 4558.38849$\pm$0.00173 & 86.60$\pm$0.24 & 0.466$\pm$0.016 & 3.99$\pm$0.33 & 6.60$\pm$0.87 & 66.6$\pm$5.6\\
129 & NMSU 1m & V & 4563.67937$\pm$0.00257 & 86.45$\pm$0.33 & 0.467$\pm$0.019 & 3.87$\pm$0.61 & 6.34$\pm$1.12 & 61.5$\pm$8.0\\
129 & APO 3.5m & V & 4563.67968$\pm$0.00051 & 86.44$\pm$0.17 & 0.459$\pm$0.015 & 4.73$\pm$0.28 & 8.60$\pm$0.44 & 61.8$\pm$2.2\\
132 & James Roe\tablenotemark{a}  & B & 4571.61844$\pm$0.00107 & 88.60$\pm$0.62 & 0.455$\pm$0.015 & 5.24$\pm$0.24 & 14.84$\pm$0.83 & 95.5$\pm$3.6\\
137 & NMSU 1m & I & 4584.83301$\pm$0.00117 & 86.55$\pm$0.19 & 0.449$\pm$0.015 & 5.89$\pm$0.51 & 14.61$\pm$1.72 & 65.2$\pm$3.6\\
137 & APO 3.5m & V & 4584.83084$\pm$0.00035 & 86.32$\pm$0.16 & 0.464$\pm$0.015 & 4.20$\pm$0.20 & 6.36$\pm$0.25 & 58.3$\pm$1.2\\
137 & SBO 24$"$ & I & 4584.82868$\pm$0.00166 & 86.63$\pm$0.21 & 0.448$\pm$0.015 & 5.95$\pm$0.50 & 15.23$\pm$2.13 & 67.3$\pm$3.8\\
137 & Bruce Gary\tablenotemark{a} & R & 4584.82876$\pm$0.00087 & 86.51$\pm$0.18 & 0.463$\pm$0.015 & 4.32$\pm$0.25 & 7.41$\pm$0.58 & 64.0$\pm$2.4\\
138 & Manuel Mendez\tablenotemark{a} & R & 4587.47754$\pm$0.00170 & 86.91$\pm$0.28 & 0.462$\pm$0.016 & 4.45$\pm$0.35 & 8.83$\pm$1.13 & 73.9$\pm$4.9\\
140 & CC 11$"$ & None & 4592.76123$\pm$0.00140 & 86.25$\pm$0.17 & 0.463$\pm$0.015 & 4.38$\pm$0.48 & 6.64$\pm$1.03 & 56.0$\pm$3.4\\
140 & APO 3.5m & V & 4592.76281$\pm$0.00084 & 86.50$\pm$0.17 & 0.465$\pm$0.015 & 4.12$\pm$0.21 & 6.71$\pm$0.36 & 63.5$\pm$3.4\\
140 & SBO 24$"$ & I & 4592.76202$\pm$0.00177 & 86.55$\pm$0.26 & 0.453$\pm$0.017 & 5.43$\pm$0.84 & 12.25$\pm$1.50 & 65.3$\pm$4.8\\
143 & SBO 24$"$ & I & 4600.69795$\pm$0.00118 & 85.88$\pm$0.24 & 0.425$\pm$0.067 & 8.52$\pm$6.54 & 6.75$\pm$1.08 & 42.0$\pm$8.5\\
146 & James Roe\tablenotemark{a}  & V & 4608.62470$\pm$0.00107 & 86.32$\pm$0.23 & 0.454$\pm$0.015 & 5.31$\pm$0.67 & 9.86$\pm$0.55 & 58.4$\pm$6.3\\
\nodata & 3.5m Data Combined & V & \nodata & 86.39$\pm$0.16 & 0.463$\pm$0.015 & 4.39$\pm$0.22 & 7.25$\pm$0.31 & 60.6$\pm$1.3\\
\cutinhead{Star and Planet Radii Fixed by Fixing k}
-318 & NMSU 1m & V & 3381.85596$\pm$0.00212 & 86.15$\pm$0.17 & 0.464$\pm$0.016 & 4.23$\pm$0.28 & 5.61$\pm$0.63 & 52.5$\pm$3.2\\
0 & \citet{Gillon07b}\tablenotemark{b} & V & 4222.61617$\pm$0.00062 & 86.40$\pm$0.16 & 0.464$\pm$0.016 & 4.23$\pm$0.28 & 6.74$\pm$0.69 & 60.5$\pm$2.6\\
1 & \citet{Shporer08}\tablenotemark{b} & None & 4225.26049$\pm$0.00094 & 86.45$\pm$0.19 & 0.464$\pm$0.016 & 4.23$\pm$0.30 & 6.86$\pm$0.76 & 61.8$\pm$3.7\\
1 & \citet{Shporer08}\tablenotemark{b} & V & 4225.26050$\pm$0.00076 & 86.39$\pm$0.16 & 0.464$\pm$0.016 & 4.23$\pm$0.30 & 6.65$\pm$0.72 & 59.9$\pm$2.1\\
9 & \citet{Shporer08}\tablenotemark{b} & R & 4246.41009$\pm$0.00103 & 86.37$\pm$0.17 & 0.464$\pm$0.017 & 4.23$\pm$0.29 & 6.66$\pm$0.70 & 59.6$\pm$3.6\\
22 & \citet{Gillon07a}  & 8$\micron$ & 4280.78219$\pm$0.00011 & 86.34$\pm$0.16 & 0.464$\pm$0.016 & 4.23$\pm$0.30 & 7.47$\pm$1.00 & 59.5$\pm$1.0\\
110 & Gregor Srdoc\tablenotemark{a} & R & 4513.43416$\pm$0.00191 & 86.19$\pm$0.18 & 0.464$\pm$0.016 & 4.23$\pm$0.28 & 5.92$\pm$0.74 & 54.0$\pm$4.3\\
110 & Tonny Vanmunster\tablenotemark{a} & R & 4513.44424$\pm$0.00386 & 87.29$\pm$0.54 & 0.464$\pm$0.016 & 4.23$\pm$0.28 & 8.20$\pm$1.11 & 77.5$\pm$9.0\\
112 & Bruce Gary\tablenotemark{a} & R & 4518.73038$\pm$0.00358 & 86.20$\pm$0.25 & 0.464$\pm$0.017 & 4.23$\pm$0.31 & 6.00$\pm$1.03 & 55.7$\pm$7.3\\
113 & Gregor Srdoc\tablenotemark{a}  & R & 4521.37312$\pm$0.00244 & 87.34$\pm$0.39 & 0.464$\pm$0.016 & 4.23$\pm$0.31 & 8.32$\pm$1.26 & 80.1$\pm$6.3\\
115 & James Roe\tablenotemark{a} & V & 4526.66055$\pm$0.00302 & 86.14$\pm$0.27 & 0.464$\pm$0.015 & 4.23$\pm$0.28 & 5.56$\pm$1.30 & 51.7$\pm$9.1\\
115 & Joao Gregorio\tablenotemark{a} & V & 4526.65996$\pm$0.00101 & 86.65$\pm$0.19 & 0.464$\pm$0.016 & 4.23$\pm$0.29 & 7.42$\pm$0.93 & 67.3$\pm$3.4\\
117 & Richard Schwartz\tablenotemark{a} & V & 4531.94392$\pm$0.00198 & 86.26$\pm$0.26 & 0.464$\pm$0.016 & 4.23$\pm$0.31 & 5.95$\pm$1.08 & 56.1$\pm$7.5\\
118 & \citet{Alonso08} & H & 4534.59610$\pm$0.00014 & 86.40$\pm$0.16 & 0.464$\pm$0.016 & 4.23$\pm$0.28 & 7.42$\pm$0.98 & 61.1$\pm$1.0\\
127 & Manuel Mendez\tablenotemark{a} & R & 4558.38809$\pm$0.00164 & 86.51$\pm$0.21 & 0.464$\pm$0.016 & 4.23$\pm$0.23 & 7.00$\pm$0.76 & 64.2$\pm$4.7\\
129 & NMSU 1m & V & 4563.67966$\pm$0.00252 & 86.38$\pm$0.23 & 0.464$\pm$0.016 & 4.23$\pm$0.28 & 6.63$\pm$0.79 & 60.1$\pm$5.3\\
129 & APO 3.5m & V & 4563.67971$\pm$0.00116 & 86.53$\pm$0.18 & 0.464$\pm$0.016 & 4.23$\pm$0.28 & 7.06$\pm$0.78 & 64.1$\pm$3.2\\
132 & James Roe\tablenotemark{a} & B & 4571.61831$\pm$0.00467 & 88.62$\pm$1.02 & 0.464$\pm$0.016 & 4.23$\pm$0.28 & 9.25$\pm$1.27 & 93.3$\pm$6.5\\
137 & NMSU 1m & I & 4584.83373$\pm$0.00379 & 86.84$\pm$0.78 & 0.464$\pm$0.016 & 4.23$\pm$0.27 & 7.70$\pm$1.19 & 72.6$\pm$14.5\\
137 & APO 3.5m & V & 4584.83084$\pm$0.00036 & 86.32$\pm$0.16 & 0.464$\pm$0.016 & 4.23$\pm$0.29 & 6.45$\pm$0.62 & 58.0$\pm$1.8\\
137 & SBO 24$"$ & I & 4584.82787$\pm$0.00912 & 86.64$\pm$0.73 & 0.464$\pm$0.016 & 4.23$\pm$0.28 & 7.05$\pm$1.45 & 67.8$\pm$17.1\\
137 & Bruce Gary\tablenotemark{a} & R & 4584.82874$\pm$0.00100 & 86.53$\pm$0.18 & 0.464$\pm$0.016 & 4.23$\pm$0.28 & 7.07$\pm$0.76 & 64.3$\pm$2.8\\
138 & Manuel Mendez\tablenotemark{a} & R & 4587.47761$\pm$0.00204 & 87.00$\pm$0.31 & 0.464$\pm$0.016 & 4.23$\pm$0.29 & 7.92$\pm$0.99 & 75.0$\pm$5.3\\
140 & CC 11$"$ & None & 4592.76119$\pm$0.00142 & 86.27$\pm$0.15 & 0.464$\pm$0.016 & 4.23$\pm$0.29 & 6.24$\pm$0.66 & 56.5$\pm$2.9\\
140 & APO 3.5m & V & 4592.76248$\pm$0.00093 & 86.47$\pm$0.17 & 0.464$\pm$0.016 & 4.23$\pm$0.29 & 6.94$\pm$0.68 & 62.6$\pm$2.9\\
140 & SBO 24$"$ & I & 4592.76090$\pm$0.00430 & 86.71$\pm$0.32 & 0.464$\pm$0.016 & 4.23$\pm$0.28 & 7.57$\pm$1.06 & 69.5$\pm$7.8\\
143 & SBO 24$"$ & I & 4600.69668$\pm$0.00171 & 86.08$\pm$0.17 & 0.464$\pm$0.016 & 4.23$\pm$0.30 & 5.54$\pm$0.71 & 50.4$\pm$3.6\\
146 & James Roe\tablenotemark{a} & V & 4608.62542$\pm$0.00508 & 86.40$\pm$0.68 & 0.464$\pm$0.016 & 4.23$\pm$0.28 & 6.53$\pm$2.03 & 63.8$\pm$19.6\\
\nodata & 3.5m Data Combined & V & \nodata & 86.43$\pm$0.16 & 0.464$\pm$0.016 & 4.23$\pm$0.28 & 6.82$\pm$0.67 & 61.7$\pm$2.7\\
\enddata
\tablenotetext{a}{Amateur Observer with data obtained from Bruce Gary. \url{http://brucegary.net/AXA/GJ436/gj436.htm}}
\tablenotetext{b}{Data were digitized from published plot}
\tablenotetext{c}{Johnson-Cousins System}
\tablecomments{All errors are 1$\sigma$}
\end{deluxetable}

\end{document}